# Is the Universe roughly-tuned for computing?

Zoltán Galántai
Budapest University of Technology and Economics, Hungary
galantai@finance.bme.hu

*"If you think about computing, there isn't just one way to compute, just like there's not just one way to move around." (Michael Dell)*

## Abstract
This short paper proposes an alternative theory to Anthropic Principle. According to our interpretation, the Universe is not "fine-tuned" for life, but "roughly-tuned" for computation and its biofilness is only a phenomenon. This standpoint allows us to extend Seth Lloyd's concept about the ultimate physical limits of computing to examine the computing capabilities of any imaginable universe. In addition, I draw up a universe classification based on it.

## Introduction

According to the Weak Anthropic Principle (WAP), „our location in the Universe is necessarily privileged to the extent of being compatible with our existence as observers", since an intelligent being without the special parameters of fundamental physical constants would not be able to come into existence. Anthropic Principle's strong version (SAP) states that this fine-tuning is not an accident. The Universe necessarily able to „admit the creation of observers within it at some stage." [Barrow - Tipler, 1986, 1 - 2]

There are several answers for either WAP or SAP. We can accept one or another form of them as a legitimate description of physical reality. Or, we can refuse them (i.e. arguing that they are based on coincidences). Nevertheless, we can propose an alternative hypothesis, i.e. that our Universe is not "fine-tuned" for life, but "roughly-tuned" for computation.

Let's clarify our topic at this point.

The following is not about the existence or nonexistence of other universes (perhaps with different constants). Until now there is no adequate theoretical or observational support for their existence, so it is only a thought-experiment about some theoretical possibilities in connection with computing.

On the other hand, Ken Wharton has sound arguments against the belief "that the universe must compute itself in the same manner" as a real computer [Wharton, 2012]. Agreeing with him, I am only to inquire into the theoretical possibilities of computing influenced by given physical parameters.

Finally, we shall ignore today's silicon based computers' problems which are outcomes of actual technological problems. Their limitations determined by the laws of physics, but these laws define only the boundaries of the technological possibilities' phase space.

We shall focus solely on theoretical questions and limits caused by some fundamental features of nature, as Shannon did it in his seminal treatise on information, which can be interpreted as an examination of theoretical limits of the amount of information transmitted via a noisy cannel [Shannon - Weaver, 1949].

Bennett and Landauer [1985] extended this kind of abstract approach writing about some fundamental physical limits of information processing, and hypercomputation makes a further step to theoretical "Platonist computers" which are incomparable to Turing computability. This presupposes in one form or other the existence of actual infinite in the physical world, and the Malament – Hogarth universes is an attempt to build a model based on some special features of a relativistic spacetime, where the difference between computable and uncomputable operations disappear [Barrow 2007, 31 - 32].

And there are other, at least imaginable solutions to exceed the limits determined by our physics in computation.

## Computing with different parameters or constants

Seth Lloyd pointed out that the maximum computing capacity of a computer as a physical system is limited by the Planck constant ($\hbar$); the speed of light (c); the Boltzmann constant ($k_B$) and the gravitational constant G. Lloyd's "Ultimate Laptop" with one kg mass and one liter volume can perform $5.4258 \times 10^{50}$ logical operations on about $10^{31}$ bits. *Ad analogiam*, knowing the mass of our Universe, we can calculate its maximum computing capacity [Lloyd, 1999] and we can play with the idea of possibilities limited by different constants than ours'.

According to the supporters of the Anthropic Principle, life is strictly bounded to particular parameters of physical constants, and a small change can be fatal. The classic example is that a strong nuclear force 2% stronger would change the formation of helium, and this would presumably prevent the formation of stars as we know them. So Earthlike life would be impossible [Davies 1993, 70 - 71].

On the other hand, we can perform different amount of operations for a given unit of time in a universe where the parameters of the four above mentioned constants are different. A higher value for the speed of light results bigger computing capacity, a bigger gravitational constant causes a smaller one etc. (note that it is still not known whether the laws and constants are mutually independent of each other). Although we can change the parameters determining the efficiency of computing, several different "computable universes" possible (with higher or lower level speed of light etc.). The computation function is not sensitive to changes. Ed Fredkin examined the problem of "missing workload" in digital cosmology, namely, that the computational capacity of the space-time is in his model is mainly idle [Fredkin, no date]. Our problem is rather different, since our Universe seems to be not idle, but to be an inefficient computing mechanism. But universes with different types of constants can lead to both different and more efficient types of computing, although in other cases they would be less effective.

We can catalogue them on the basis of their relations to the constants as it is presented in Table 1.

*Table 1. Possible types of computing*

| type | description |
|------|-------------|
| C1 | normal computing (in our Universe) |
| C2 | computing based on different parameters of our constants |
| C3 | computing based on different constants - in some cases it is impossible to interpret Lloyd's equations for them |

It have to be underlined that although the meaning of Universes with "different types of constants" is ill-defined, if defined at all, we can distinguish two different types of them.

The first one is where some constants are missing, while other constants, which are present in our Universe, exist there. For example, without photons the meaning of speed of light is uninterpretable, and a further question whether the lack of speed of light as a boundary speed means that there is no boundary of speed at all, or a universe with boundary of speed based on other components.

In the second type of universes, there are totally different forces and laws with totally different constants.

## Roughly-tuned for computing?

As we have seen, life seems to be very sensitive even for small changes of certain parameters, but computer-friendliness is resistant to them. To top it all, changes in other constants' values doesn't have any influence on computer-friendliness.

From Anthropic Principle's point of view, our Universe's computing-friendliness is only a required condition. A universe without some causality and some kinds of conversation laws is not only uninterpretable as a computable system, but it wouldn't be able to produce living beings at all (this problem was explained, inter alia, by Max Tegmark [1997]).

Opposing to this approach, it can be argued that our Universe is not fine-tuned for life, but roughly-tuned for computation, since this is its fundamental feature, and a universe which is unsuitable for life, can remain roughly-tuned for computing with a ranges of possible parameters and constants (although not with any of them).

So since computing-friendliness seems to be much less depending on the concrete physical features of a universe than biofilness, this can be regarded as a more typical feature than life. On the contrary, biofilness is simply an epiphenomenon.

## Classification of possible Universes from computational point of view

Tegmark introduced the idea of a "mathematical democracy" where "universes governed by other equations are equally real". Its phase space contains every possible mathematics [Tegmark, 2005]. But according to Landauer, the platonic realm of mathematics and the real Universe are different, since the latter is limited by some physical restrictions. There are mathematical operations which due to the physical limitations, theoretically cannot be performed [Landauer, 1986].

Landauer focused on the possibilities allowed by our physics, but as we have seen, we can ask questions about the physical limitations of other universes. In other words, we can describe universes regarding their computer-friendliness as shown in Table 2 (these categories not necessarily overlaps our C1 - C3 classifications in Table 1).

*Table 2. Types of computational universes based on the speed of light*

| type | speed of light | C1 | C2 | C3 |
|---|---|---|---|---|
| U1 normal | $0 < c < \infty$ | x | x | x |
| U2 hypercomputable | $c = \infty$ | --- | x | x |
| U3 zero | $c = 0$ | --- | x | x |

Table 2 is based on the speed of light as an example. We can change other constants' parameters which are fundamental for computing, as well. What if G's numerical part is not $6.67384 \times 10^{-11}$, and not 1 or 2 or any other, but either 0 or infinite? etc.

**U1** is a universe with ordinary constants, with either the same or different parameters - its computing can be either faster or slower than in our world, but it is essentially the same.

**U2** is a kind of a Universe which can perform even infinite calculation within a finite amount of time, since the speed of light is not a barrier, and thus, hypercomputing is a physical reality (notice that we don't make any distinction between countably and unaccountably infinite sequences, although this can lead to further questions).

Regarding the transmission of information, the situation in a U2 world can be comparable with Newton's action at distance concept, where an object can affect another object without time, in a nonlocal way.

**U3** is a Universe without any possibility for computing: e.g. since *c* as a maximum speed is zero, there is no information transmission at all.

## Life, universe and computing

At least one U1 universe is biofil, and the question about fine-tuning can be interpreted as a question whether the result of other combinations of our constants' parameters can result habitable environments.

Anthropic Principle suggests that the answer is a clear "no", but according to another opinion a *ceteris paribus*: changing only one parameter one time approach is misleading in this case. If you choose smaller tires to your car, they wouldn't fit on the wheels, but you can buy another car with different wheel size [Cohen - Stewart 2002, 5-7], and the situation is the same in connection with life's fine-tuning. So we have to make a distinction between fine-tuning and fragility (or probability of other forms of life fine-tuned to either different parameters or different constants), and following this logic, fine-tuning seems not to be a proof for uniqueness [ibid, 13]. This seems to be convincing, but to accept this metaphor, we have to accept that – similarly to different car brands – different types of life exist. But this is exactly the question, and nobody knows the answer.

So the most we can say is that the possibility of other kinds of life which are fine-tuned to different constants, parameters, etc. cannot be ruled out, although it is plausible that there are more combinations of parameters of constants, which can lead to computable, than a biofil solution either in a U1 or in a U2 universe.

On the other hand, if an ensemble of infinitely many universes exists, then there are equally infinitely many universes with biofil or lifeless parameters (i.e. U3 universes). Obviously it's only a possibility which cannot be excluded actually, although neither the existence of a multiverse nor even the existence of a U3 universe is not proved. *Ad absurdum*, according to our actual knowledge (or ignorance), it is imaginable that either a U3 or a computable, but lifeless universe is theoretically impossible (although it seems to be improbable).

This raises some questions. Leibniz assumed in the 17th century that there must be a reason for a thing's existence, so from his point of view it was a fundamental question why something existed instead of nothing [Holt 2013, 11].

The central problem of Anthropic Principle is why a biofil universe exists. Following this way of logic we can ask a similar question.

We can argue that it is not surprising to observe a computable universe around us, because an uncomputable one is not suitable for life. But this is not an answer for the question that why exists a U1 universe instead of either a U2 or U3 universe or instead of nothing.

And a final question: What kind of other features can exist in our Universe in addition to bio- and computation-friendliness, which are worth to examine for fine-tuning?